\documentclass{article}

\usepackage{microtype}
\usepackage{graphicx}
\usepackage{subfigure}
\usepackage{booktabs} %
\usepackage{pifont} %

\usepackage{hyperref}
\usepackage{xurl} %

\usepackage[accepted]{mlsys2025}
\makeatletter
\renewcommand{\Notice@String}{Preprint.}
\makeatother

\usepackage{kotex}

\usepackage{xspace}
\newcommand{\sysnameplain}{Bevy}
\newcommand{\sysname}{\sysnameplain\xspace}
\newcommand{\sysnamebf}{\textbf{\sysnameplain}\xspace}

\newcount\layoutcnt
\newcommand{\blanklines}[1]{%
  \par\noindent\mbox{}%
  \layoutcnt=\numexpr#1-1\relax
  \loop\ifnum\layoutcnt>0 \\\mbox{}\advance\layoutcnt by -1 \repeat
  \par}

\mlsystitlerunning{Accelerating Offloading-Based MoE Inference by Interleaving Concurrent Requests}

\begin{document}

\twocolumn[
\mlsystitle{Accelerating Offloading-Based MoE Inference \\ by Interleaving Concurrent Requests}

\mlsyssetsymbol{equal}{*}

\begin{mlsysauthorlist}
\mlsysauthor{Hongseung Yu}{snu}
\mlsysauthor{In Gim}{yale}
\mlsysauthor{Kyunghan Lee}{snu}
\mlsysauthor{Lin Zhong}{yale}
\end{mlsysauthorlist}

\mlsysaffiliation{snu}{Seoul National University}
\mlsysaffiliation{yale}{Yale University}

\mlsyscorrespondingauthor{Kyunghan Lee}{kyunghanlee@snu.ac.kr}
\mlsyscorrespondingauthor{Lin Zhong}{lin.zhong@yale.edu}

\mlsyskeywords{Machine Learning, MLSys}

\vskip 0.3in

\begin{abstract}
Expert offloading enables large mixture-of-experts (MoE) language models to run locally, but conventional batching stalls the GPU while missing experts are fetched.
A popular remedy, known as prefetching, is to predict what experts will be needed soon and fetch them using idle link bandwidth. Imprecise prediction, however, wastes bandwidth and GPU memory on fetching unnecessary experts.
Our key finding is that prefetching can be harmful when there is an expert cache in GPU memory. This is because as the cache includes more necessary experts, expert prediction becomes less precise.
Our key insight is that proper request scheduling beats prefetching.
Building on this insight, we present \sysnamebf, an MoE inference system based on runahead inference for concurrent requests.
A request whose experts are available can \emph{run ahead} to later layers while other requests wait for fetches, revealing the experts it needs next.
\sysname can then fetch the missing experts so identified while computing ready requests, overlapping computation and fetching without predicting expert selections.
Running requests separately, however, repeats reads of shared weights in GPU memory and increases compute time.
To limit this cost, \sysname's \emph{self-balancing} scheduler batches ready requests at each layer and prioritizes lagging requests, adapting their grouping as expert availability changes.
Across two local platforms, \sysname delivers up to 3.46$\times$ the decode throughput of llama.cpp and up to 1.55$\times$ that of the fastest prefetching baseline.
\end{abstract}
]

\printAffiliationsAndNotice{}  %

\section{Introduction}
\label{sec:intro}
\suppressfloats[t]

\begin{figure}[t]
\centering
\includegraphics[width=\columnwidth]{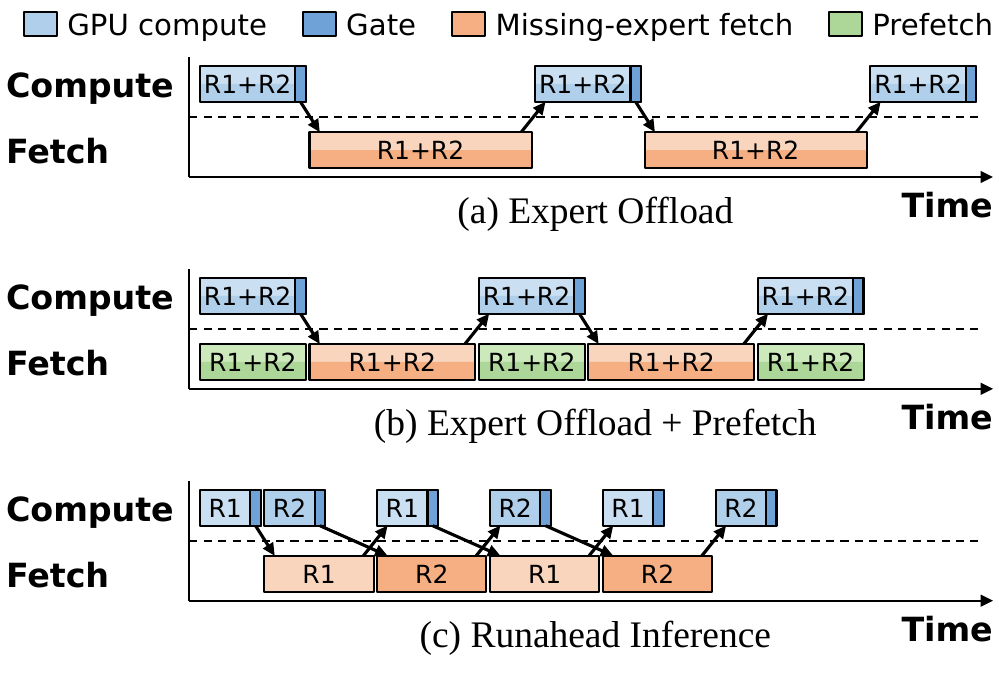}
\vspace{-5mm}
\caption{Timeline of two concurrent requests, R1 and R2. (a) With batching, the GPU idles during fetches and the link idles during computation. (b) Same as (a), but predicted experts are prefetched during computation. Because prediction is never perfect, prefetching unnecessary experts wastes both bandwidth and GPU memory. (c) Each request runs without waiting for the other's fetches, so fetches overlap with computation.}
\label{fig:timeline}
\end{figure}

Large language models (LLMs) are increasingly deployed on local systems such as personal desktops and AI workstations~\cite{macmini,rtxspark}.
Although such local deployments typically serve a single user, applications that use these models often perform multiple tasks concurrently, generating several requests to the underlying model.
Personal agents may execute subtasks in parallel, while chat applications can process multiple queries on the same model simultaneously~\cite{openclaw,lmstudio}.
The ability to serve concurrent requests efficiently is therefore becoming more important for local inference.

Local inference must make efficient use of limited compute resources.
Mixture-of-experts (MoE) models are well suited to this setting because they reduce computation by activating only a few experts per token.
Their full set of expert weights, however, can still exceed the memory available for inference on a personal device.
Expert offloading systems keep a subset of the experts in GPU memory as an expert cache, managed by a replacement policy such as LRU or LFU.
Missing experts are fetched from secondary storage over a link such as PCIe~\cite{llamacpp,ssdllama,moeinfinity}.
As shown in Figure~\ref{fig:timeline}(a), after each layer's gate network selects the experts, the system fetches those absent from the cache before computing the MoE.
This sequence repeats at every layer, leaving the GPU idle during fetches and the link idle during computation.

Prefetching addresses this GPU idle time by predicting experts needed at later layers and fetching them while the GPU computes (Figure~\ref{fig:timeline}(b))~\cite{finemoe,mixtraloffloading,promoe}.
However, we find that the benefit of prefetching is questionable when combined with an expert cache (\S\ref{sec:bm-motivation}).
The cache already holds many of the experts needed next and tends to retain those that are easier to predict.
The prefetcher must therefore choose among the remaining experts, of which needed experts are less common and harder to predict.
Prefetching can keep the link busy, but much of that bandwidth is wasted fetching experts that requests do not need.

Our key insight is that concurrent requests can use otherwise idle link bandwidth to fetch experts known to be needed, without prediction.
That is, proper request scheduling beats prefetching.
Conventional batching computes requests together at each layer after the experts needed by the entire batch have arrived.
Allowing a request to proceed as soon as its own experts are available lets it reach the next layer earlier and identify the experts needed there, including those missing from the cache.
The system can then fetch these newly identified missing experts while the GPU computes other requests whose experts have arrived (Figure~\ref{fig:timeline}(c)).
We call this execution method \emph{runahead inference}.

Building on this idea, we present \sysname, an MoE inference system for concurrent requests.
A key challenge is that overlapping computation and expert fetching does not by itself guarantee a shorter inference step.
Running requests separately repeats reads of shared weights in GPU memory, increasing compute time and potentially offsetting the benefit of overlap.
We find that statically grouping requests, e.g., based on profiling as in NanoFlow~\cite{nanoflow}, is inadequate.
This is because the number of expert cache misses, and hence the fetch time, varies substantially from step to step (Figure~\ref{fig:c1}).
Static grouping can therefore leave compute time unnecessarily high in some steps and keep ready requests waiting in others.

To address this problem, \sysname groups requests layer by layer dynamically, at runtime, according to request readiness and progress.
At each scheduling decision, it batches the ready requests at the selected layer.
\sysname gives priority to requests at earlier layers, both for computation and for expert fetching.
This helps them catch up with other requests and creates more opportunities to batch requests at the same layer.
With more expert misses, fewer requests tend to be ready together, so smaller batches run ahead while experts are fetched for other requests.
With fewer misses, more requests tend to become ready together, allowing larger batches to reduce compute time.
Together, these policies make execution \emph{self-balancing}:
grouping adapts across layers within a step as well as across steps, without profiling or a fixed partition.

We implement \sysname on llama.cpp~\cite{llamacpp} with Metal and CUDA support and evaluate it on two local inference platforms using two MoE models and chat and agentic coding workloads.
Across experiments varying concurrency and expert cache size, \sysname delivers 1.26--3.46$\times$ the decode throughput of llama.cpp and 1.01--1.55$\times$ that of the fastest prefetching baseline.
\sysname achieves these gains without predicting expert selections, by exposing demand early and adapting request grouping during execution.

\section{Background \& Motivation}
\label{sec:background}

\subsection{Concurrent Requests in Local LLM Inference}
\label{sec:bm-concurrent}

As LLMs spread to more applications, serving several requests at the same time is becoming common. LM Studio, a chat application, handles multiple queries from a user on the same model simultaneously~\cite{lmstudio}. Personal agents such as OpenClaw execute code review and research sessions concurrently while the user continues an interactive session~\cite{openclaw}. These agents also often split a single task, such as checking a mailbox, into four to eight subtasks and run them in parallel~\cite{nvidiaifa2026}. For such tasks, concurrent request throughput directly affects how long a single user waits for the overall result. Local inference thus increasingly needs to handle such concurrent requests efficiently.

\subsection{Offloading-Based MoE Inference}
\label{sec:bm-offloading}

A mixture-of-experts (MoE) LLM activates its experts sparsely: in each layer, the gate network after the attention picks a few experts for each token, so each inference step uses only a fraction of the model's weights, which reduces computation. The set of experts needed in each step varies and is not known in advance, so the model still needs all of its experts for inference. When the experts do not fit in GPU memory, expert offloading systems store all of them in secondary storage (host memory or SSD) and fetch the experts needed for computation into GPU memory over a link such as PCIe~\cite{freetoken,llamacpp,ssdllama,moeinfinity}.

The fetching strategy depends on how many tokens a step processes. Prefill of long input prompts and decode with large batches of concurrent requests process many tokens per step. Since the gate network picks a different set of experts for each token, these steps often need most of the experts in the model. The systems then fetch every expert of the next layer from secondary storage into GPU memory over the link ahead of time, in the same layerwise pipeline used to offload dense models~\cite{flexgen}. In contrast, decode with a handful of concurrent requests (\S\ref{sec:bm-concurrent}) processes few tokens per step and uses only a fraction of the experts. Since fetching whole layers is then inefficient, the systems instead keep a subset of the experts in GPU memory as an \emph{expert cache} managed by a replacement policy such as LRU or LFU. After the gate network of a layer picks its experts, they fetch only the missing experts, those not in the expert cache, on demand over the link~\cite{finemoe,freetoken,moeinfinity,ssdllama}.

\subsection{Prefetching to Use Idle Link Bandwidth}
\label{sec:bm-prefetch}

On-demand fetches delay the step, since the computation cannot proceed until all the missing experts arrive (Figure~\ref{fig:timeline}(a)). To address this problem, many expert offloading systems \emph{prefetch} experts. A prefetcher predicts the experts needed by a later layer that the GPU has not yet computed and fetches them into the expert cache in advance. As shown in Figure~\ref{fig:timeline}(b), the prefetcher fetches the experts while the GPU computes, so it also utilizes the otherwise idle link bandwidth. Prefetchers predict the experts in various ways. For instance, some feed the current hidden state into the gate network of the later layer~\cite{mixtraloffloading,moeapex}, some use past activation history and correlations~\cite{moeinfinity,finemoe,klotski}, and others train a prediction model~\cite{promoe,expertflow,expertflowshen}.

\begin{figure}[t]
\centering
\includegraphics[width=\columnwidth]{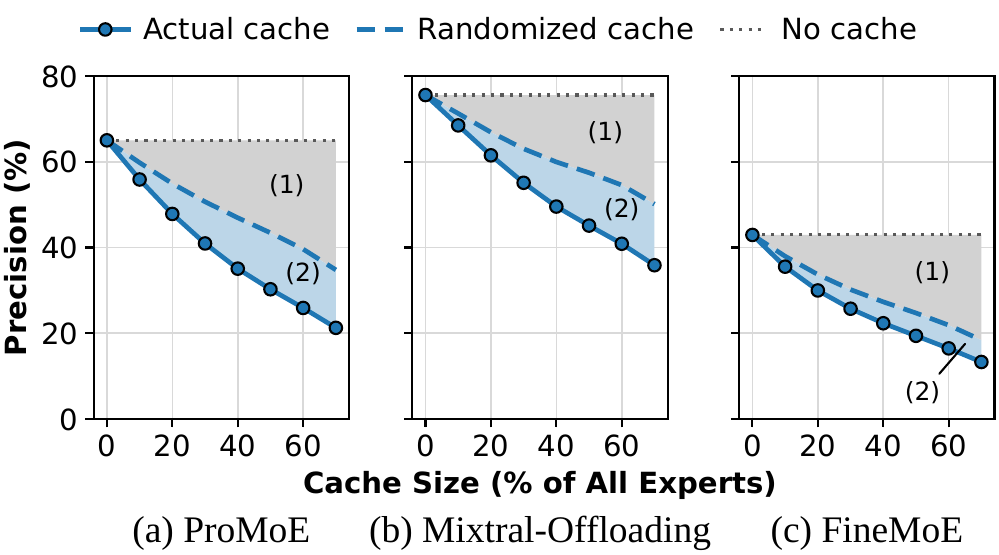}
\vspace{-7mm}
\caption{Prefetch precision under caching at 80\% coverage of nonresident target-layer demand. To test whether the smaller fraction of needed candidates alone explains the decline, we compare the actual cache (solid) with a randomized cache (dashed). For each cache snapshot, the randomized cache holds the same numbers of needed and unneeded experts as the actual cache, selected randomly within each group. The dotted line is the no-cache reference. Region~(1) shows the precision drop from this reference to the randomized cache; region~(2) shows the additional drop to the actual cache.
}
\label{fig:prefetch-goodput}
\end{figure}

\subsection{Caching Reduces Prefetch Precision}
\label{sec:bm-motivation}

However, we find that prefetching uses this otherwise idle bandwidth inefficiently when combined with an expert cache.

We evaluate ProMoE~\cite{promoe}, Mixtral-Offloading~\cite{mixtraloffloading}, and FineMoE~\cite{finemoe}, one prefetcher for each prediction approach above. We run them in a trace-driven cache simulator on Qwen3.6-35B-A3B~\cite{qwen36} decode traces of eight concurrent requests from ShareGPT~\cite{sharegpt}. All three prefetchers use the same LFU cache policy. We vary the cache size and compare precision---the fraction of selected nonresident experts needed by the target layer---at a fixed coverage of nonresident demand (Figure~\ref{fig:prefetch-goodput}).

Precision decreases as the cache grows for all three prefetchers. The cache already holds a large share of upcoming demand: a cache holding 30\% of all experts contains 54.1\% of the experts needed by the target layer on average. Needed experts therefore make up a smaller fraction of the nonresident candidates. To see whether this alone explains the decline, we randomize which experts are cached while keeping the fraction of needed candidates unchanged. The randomized cache reproduces part of the decline (region~1), while the actual cache shows an additional drop (region~2). This suggests that the cache tends to hold experts the prefetchers can predict accurately, so the remaining candidates are harder to predict.

Prefetching can thus keep the link busy, but much of what it fetches is not needed by the target layer. Filling idle periods with predictions therefore does not ensure that the bandwidth serves actual demand. Can we reduce idle link time while fetching only experts known to be needed? The next section presents a request scheduling method designed to achieve this goal.

\providecommand{\stepmark}[1]{\ding{\the\numexpr181+#1\relax}}

\begin{figure}[t]
\centering
\includegraphics[width=\columnwidth]{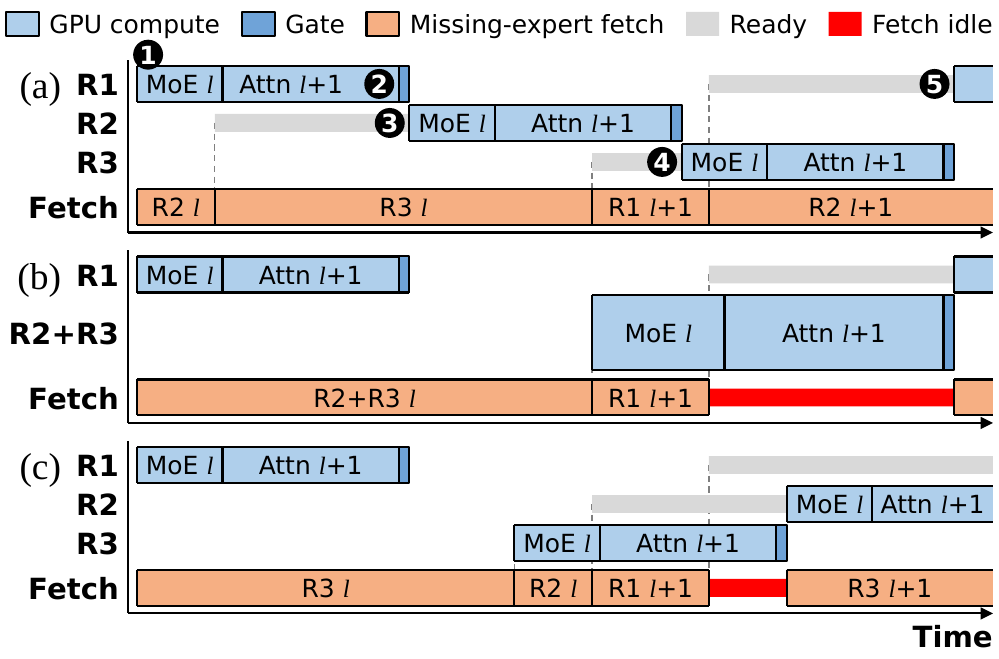}
\vspace{-7mm}
\caption{Example timeline of runahead inference with three requests that reach layer~$l$ together. (a) Each request runs without waiting for the other requests' fetches. (b) Same as (a), but R2 and R3 are grouped. (c) Same as (a), but the experts of R3 are fetched before those of R2.}
\vspace{-7mm}
\label{fig:runahead}
\end{figure}

\section{Runahead Inference}
\label{sec:idea}

In this section, we present \emph{runahead inference}, a novel scheduling method for offloading-based MoE inference. The key idea is to let ready requests \emph{run ahead} to later layers and reveal their expert needs early, without waiting for other requests' fetches. This allows the link to stay busy fetching only experts known to be needed, without relying on predictions.

\subsection{Running Requests at Different Layers} %

Figure~\ref{fig:runahead}(a) shows in detail how runahead inference runs three requests (R1, R2, and R3) that reach layer $l$ together. \stepmark{1} Only R1 finds all of its experts for layer $l$ in the expert cache, so it runs the MoE right away instead of waiting for the fetches of R2 and R3. \stepmark{2} R1 then runs the attention and the gate of layer $l{+}1$, as their weights remain in GPU memory. The gate picks its experts while the link is still busy, so the fetches of its missing experts queue behind those of R2 and R3. \stepmark{3} Meanwhile, the GPU runs R2 once its fetches finish, \stepmark{4} followed by R3. \stepmark{5} By the time R3 finishes, the fetches of R1 for layer $l{+}1$ are done, so the GPU moves on to R1 right away. As the requests proceed through the remaining layers in this way, the link bandwidth is filled only with fetches of missing experts.

\begin{figure}[t]
\centering
\includegraphics[width=\columnwidth]{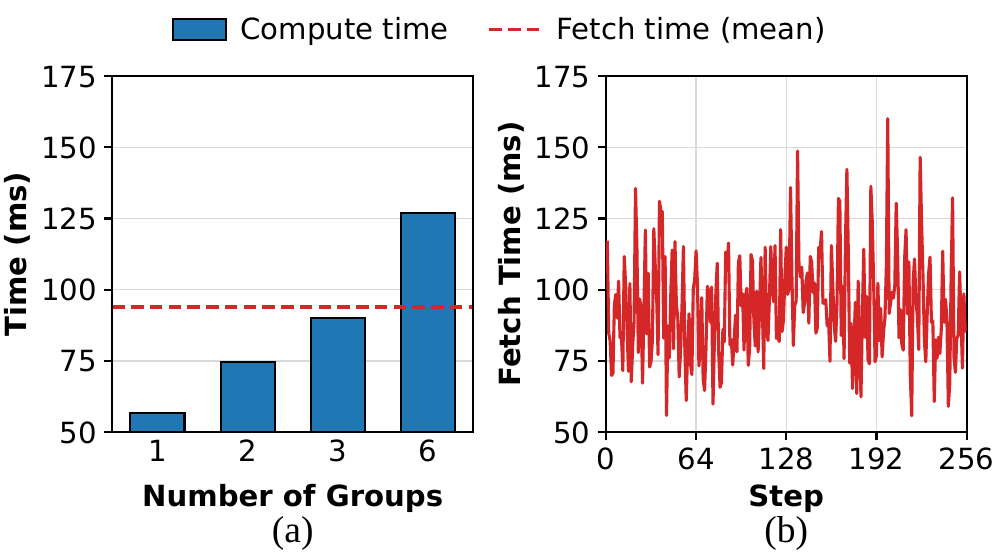}
\vspace{-7mm}
\caption{Compute and fetch time per step for six concurrent requests (Gemma4-26B-A4B, Agentic Coding, 30\% expert cache, Apple M6). (a) Compute time when the requests run in 1, 2, 3, or 6 groups. (b) Fetch time of each step over 256 steps.}
\label{fig:c1}
\end{figure}

\subsection{Challenges in Grouping and Ordering}
\label{sec:idea-challenges}

To understand when runahead inference works well, we first define the compute time and fetch time of a step. The \emph{compute time} is the total time the GPU is busy computing during the step. For the three requests in Figure~\ref{fig:runahead}(a), it is the sum of the time the GPU spends on R1, R2, and R3. Similarly, the \emph{fetch time} is the total time the link is busy fetching the step's missing experts. A step must complete both computation and fetching, so it cannot be shorter than the longer of these two times. Its actual duration also depends on how much computation and fetching overlap across requests. Effective scheduling must therefore aim to (i) \emph{keep both compute time and fetch time short} and (ii) \emph{overlap computation and fetching} across requests as much as possible. However, two challenges make this difficult.

\noindent\textbf{C1: Grouping Granularity.}
Running each request separately increases compute time. With a small number of concurrent requests, GPU computation is memory-bound, so the GPU reads shared weights from GPU memory again for each request. In Figure~\ref{fig:c1}(a), computing six requests as one batch takes 56.5\,ms per step, whereas computing them one at a time takes 126.8\,ms in total, above the mean fetch time (dashed line). However much computation and fetching overlap, the step still cannot be shorter than this compute time.

One possible solution is \emph{static grouping}, which splits the requests into groups in advance and runs each group as one batch for the whole step.
For example, forming two groups of three requests or three groups of two requests reduces the compute time from 126.8\,ms to 74.6\,ms or 90.1\,ms, respectively, both below the mean fetch time (Figure~\ref{fig:c1}(a)). Static grouping, however, has two limitations. First, it restricts when requests can run ahead. In Figure~\ref{fig:runahead}(b), grouping R2 and R3 reduces the number of scheduling units from three to two. R2 must wait for R3's experts instead of computing while they are fetched. After R1's fetches for layer~$l{+}1$ finish, the rest of the group's computation runs without any overlapping fetches. Grouping can therefore reduce compute time at the cost of fewer opportunities to overlap computation and fetching.

\begin{figure}[t]
\centering
\includegraphics[width=\columnwidth]{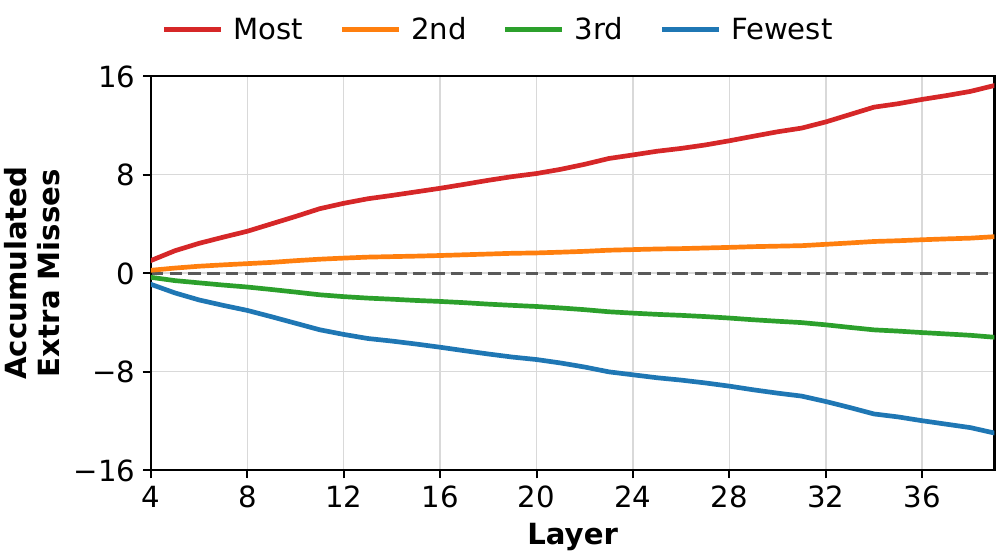}
\vspace{-7mm}
\caption{Accumulated extra misses over the mean for four concurrent requests ranked by their misses in layers 0--3 (Qwen3.6-35B-A3B, ShareGPT, 30\% expert cache).}
\label{fig:c2}
\end{figure}

Second, even under the same grouping, the fetch time and the compute time change from step to step. Because each step misses a different number of experts, the fetch time ranges from 55.7 to 160.1\,ms over 256 steps (Figure~\ref{fig:c1}(b)). The compute time also grows as the context of each request gets longer. The compute time of a grouping can thus be shorter than the fetch time in one step but longer in another, so no grouping fixed in advance suits every step. In \S\ref{sec:design-merge}, we present a mechanism that groups requests at each layer at runtime.

\noindent\textbf{C2: Scheduling Order.}
When multiple requests are ready for computation or have experts to fetch, the scheduling order affects how much computation and fetching can overlap. Figure~\ref{fig:runahead}(c) shows the same three requests as Figure~\ref{fig:runahead}(a), but the link fetches the experts of R3, which misses more, before those of R2. When R1 finishes, neither R2 nor R3 is ready, so the GPU sits idle while the link continues fetching. By contrast, in Figure~\ref{fig:runahead}(a), the link serves R2 first, so R2 is ready when R1 finishes and its computation overlaps with the remaining fetches. A natural approach is therefore to fetch first for the request with the fewest remaining missing experts, allowing its computation to start sooner and overlap with fetches for other requests.

However, always prioritizing requests with fewer misses can leave other requests behind. Figure~\ref{fig:c2} ranks four requests by their misses in layers 0--3 and tracks their accumulated extra misses over the mean in later layers. Requests that miss more early tend to keep missing more. Repeatedly favoring requests with fewer misses can therefore cause a request with more misses to accumulate delays and fall behind as a \emph{straggler}. The step cannot end until the straggler finishes its remaining fetches and computation alone, without other requests' work to overlap with. The scheduling order must therefore consider both the overlap available now and the progress of all requests through the step.

\section{\texorpdfstring{\sysname}{\sysnameplain} System Design}
\label{sec:design}

In this section, we present \sysnamebf, an MoE inference system based on runahead inference for concurrent requests.

\begin{figure}[t]
\centering
\includegraphics[width=\columnwidth]{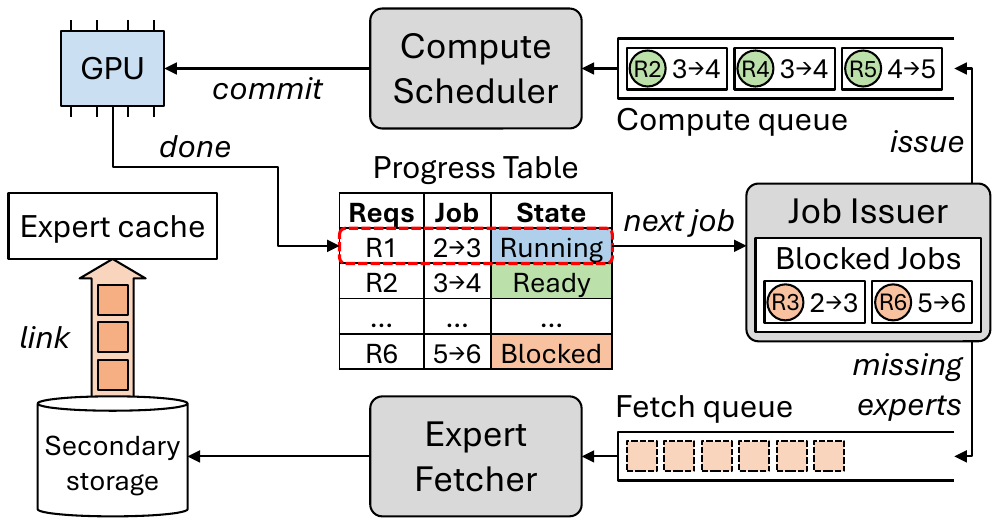}
\caption{System overview of \sysname.}
\label{fig:overview}
\end{figure}

\subsection{Overview}
\label{sec:design-overview}

Figure~\ref{fig:overview} presents an overview of \sysname, which consists of three modules: the Compute Scheduler, the Job Issuer, and the Expert Fetcher. \sysname divides the computation of each request into a sequence of jobs. Job~$l$ runs the MoE of layer~$l$, followed by the attention and the gate of layer~$l{+}1$ (Figure~\ref{fig:overview} labels it $l{\rightarrow}l{+}1$). \sysname keeps a Progress Table that records the current job and state of each request. Using this per-request information, \sysname schedules requests independently, allowing ready requests to run ahead while others wait for experts.

At the start of each step, all requests run the attention and the gate of the first layer together, then move on to their first jobs. For each job, the Job Issuer checks whether its experts are in the expert cache. If all are present, it issues the job to the compute queue. Otherwise, it blocks the job and puts its missing experts into the fetch queue. The Expert Fetcher fetches these experts from secondary storage into the expert cache over the link, and the Job Issuer issues the job once all its missing experts have arrived. The Compute Scheduler commits jobs from the compute queue to keep the GPU busy. When a job ends, the next job of the same request is passed to the Job Issuer, and this process repeats. The last job runs only the MoE of the final layer. Once all requests finish their last jobs, they run the LM head together to complete the step.

The rest of this section presents the techniques that \sysname builds on this structure to address the two challenges in \S\ref{sec:idea-challenges}. \S\ref{sec:design-priority} describes how the Compute Scheduler and the Expert Fetcher determine the order of computation and fetching (C2). \S\ref{sec:design-merge} describes how \sysname groups ready jobs at the same layer and runs them as one batch (C1). Finally, \S\ref{sec:design-cache} describes how \sysname manages the expert cache.

\begin{algorithm}[t]
\caption{\sysnameplain's Scheduling Policy}
\label{alg:select}
\begin{tabular}{@{}ll@{}}
$J_B$  & Blocked jobs with remaining missing experts \\
$J_R$  & Ready jobs \\
$l(j)$ & Layer of job $j$ \\
$m(j)$ & Number of remaining missing experts of job $j$ \\
\end{tabular}
\begin{algorithmic}
\FUNCTION{\textsc{SelectFetchJob}($J_B$) \hfill $\triangleright$ Fetcher}
\SHORTIF{$J_B = \emptyset$}{\textbf{return} none}
\STATE \textbf{return} $\arg\min_{j \in J_B} \big(l(j), m(j)\big)$ \hfill $\triangleright$ lowest-layer-first, then SRTF
\ENDFUNCTION
\FUNCTION{\textsc{SelectComputeGroup}($J_R$) \hfill $\triangleright$ Scheduler}
\SHORTIF{$J_R = \emptyset$}{\textbf{return} $\emptyset$}
\STATE $l_{\min} \leftarrow \min_{j \in J_R} l(j)$
\STATE \textbf{return} $\{\, j \in J_R \mid l(j) = l_{\min} \,\}$ \hfill $\triangleright$ one batch (\S\ref{sec:design-merge})
\ENDFUNCTION
\end{algorithmic}
\end{algorithm}

\begin{figure*}[t]
\centering
\includegraphics[width=\textwidth]{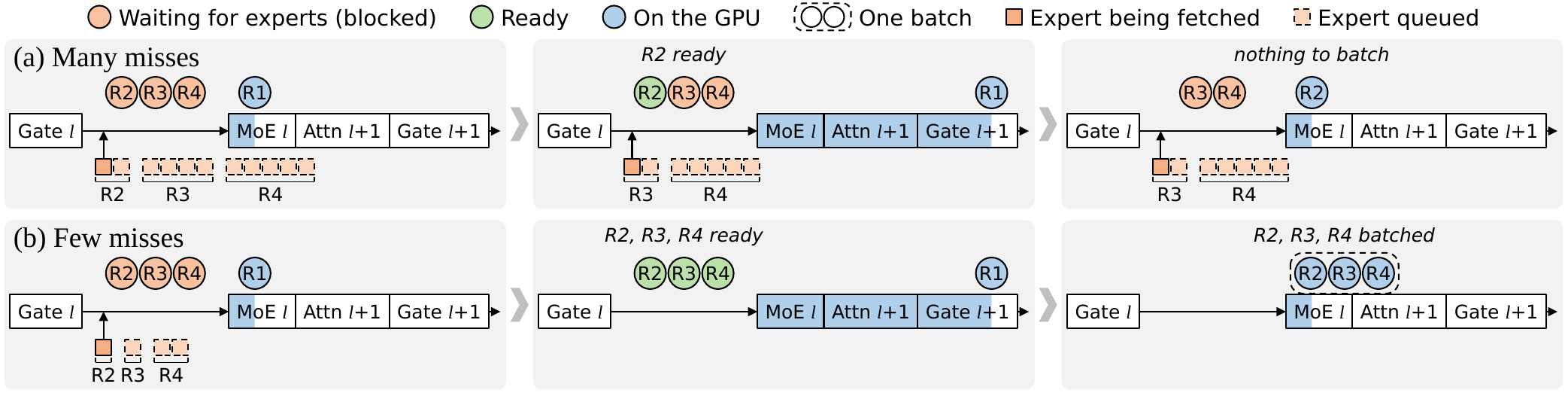}
\vspace{-5mm}
\caption{Operation of opportunistic layerwise grouping. (a) With many misses, only R2 is ready when R1 finishes, so R2 runs alone. (b) With few misses, R2, R3, and R4 are all ready and run together as one batch.}
\label{fig:merge}
\end{figure*}

\subsection{Lowest-Layer-First Scheduling}
\label{sec:design-priority}

To address C2 (\S\ref{sec:idea-challenges}), \sysname prioritizes both computation and expert fetches by request progress (Algorithm~\ref{alg:select}). It gives priority to the job at the lowest layer, that is, to the request with the least progress. This request has the most jobs left, and the step ends only when the last request finishes. Serving it first follows critical-path scheduling~\cite{hu1961} and cuts the extra waiting time of the requests that lag behind. The Scheduler selects the ready job at the lowest layer, and ready jobs at other layers run only while the lowest-layer job is blocked. The Fetcher likewise fetches the missing experts of the lowest-layer job first. This order also helps lagging requests catch up, creating batching opportunities, which we detail in \S\ref{sec:design-merge}.

Among requests at the same layer, the Fetcher first serves the one with the fewest remaining missing experts, as in shortest-remaining-time-first (SRTF) scheduling. That request becomes ready first, so its computation can overlap with the fetches of the others. Because SRTF orders only the requests at the same layer, the new fetches of a leading request never go before the pending fetches of the requests that lag behind.

\subsection{Opportunistic Layerwise Grouping}
\label{sec:design-merge}

To address C1 (\S\ref{sec:idea-challenges}), \sysname groups all ready jobs at the layer selected by the Scheduler and commits them as one batch (Algorithm~\ref{alg:select}). The group lasts only for this layer, rather than the whole step. All requests selected for a step still start and finish it together. This preserves the step-level interface used by iteration-level scheduling~\cite{orca}, allowing the request set to change between steps.

Figure~\ref{fig:merge} illustrates how the batches depend on expert misses. In both rows, R1 runs job~$l$ while the link fetches the missing experts of R2, R3, and R4 in that order (\S\ref{sec:design-priority}). With many misses (Figure~\ref{fig:merge}(a)), only R2 is ready when R1 finishes, so R2 runs alone. With few misses (Figure~\ref{fig:merge}(b)), R2, R3, and R4 are all ready, so the Scheduler batches their jobs and runs them together.

This gives layerwise grouping a self-balancing property that addresses the two limitations of static grouping in C1 (\S\ref{sec:idea-challenges}). With few misses, more jobs tend to become ready together, so batching reduces compute time when fetch time is short. With many misses, fewer jobs are ready together, so requests can run ahead in smaller batches and overlap with the remaining fetches. The grouping adapts not only between steps but also across layers within a step, without profiling. The priorities in \S\ref{sec:design-priority} help create batching opportunities: lowest-layer-first helps lagging requests catch up, while SRTF can make additional jobs at the same layer ready before the next commit. We compare this adaptive grouping with static groupings in \S\ref{sec:eval-batching}.

Waiting for more requests to become ready could yield a larger batch and further reduce compute time. For example, in Figure~\ref{fig:merge}(a), the Scheduler could wait for R3 to become ready and batch it with R2. However, waiting also delays their next gate computation and the fetches it would trigger, potentially reducing opportunities to overlap those fetches with other requests' computation. Alternatively, if a higher-layer job is ready, the Scheduler could defer R2 and run that job first. This gives R3 time to become ready but moves the leading request further ahead, potentially losing batching opportunities at later layers. In either case, the benefit of waiting depends on how the remaining jobs proceed and cannot be guaranteed at the time of scheduling. \sysname therefore batches only the jobs already ready at each commit, without delaying the selected job for additional requests.

\subsection{Expert Cache Management}
\label{sec:design-cache}
In conventional batched execution, the system fetches the union of the missing experts of all requests at a layer. With runahead inference, jobs request experts independently and can need the same expert at different times. To avoid duplicate reads, \sysname tracks each expert's pending fetch. If an expert is already queued for fetching or being fetched for another job, the new job shares that fetch rather than issuing another.

As jobs progress independently, a fetch for one job must not evict experts still needed by another. At lookup, \sysname immediately pins the cache entry of a resident expert. For a new fetch, it first reserves and pins a cache entry, evicting an unpinned expert if necessary. Each entry's reference count tracks all jobs that need it, including those sharing a pending fetch. Each job releases its reference only after completing on the GPU, and only entries with zero references can be evicted. To prevent deadlock, \sysname enables runahead inference only when the cache has enough entries to hold every concurrent request's selected experts without relying on expert sharing. For example, 128 cache entries are sufficient for 16 concurrent requests selecting eight experts each. For Qwen3.6-35B-A3B, this is only 1.25\% of all experts.

\section{Evaluation}
\label{sec:eval}
\begin{figure*}[t]
\centering
\includegraphics[width=\textwidth]{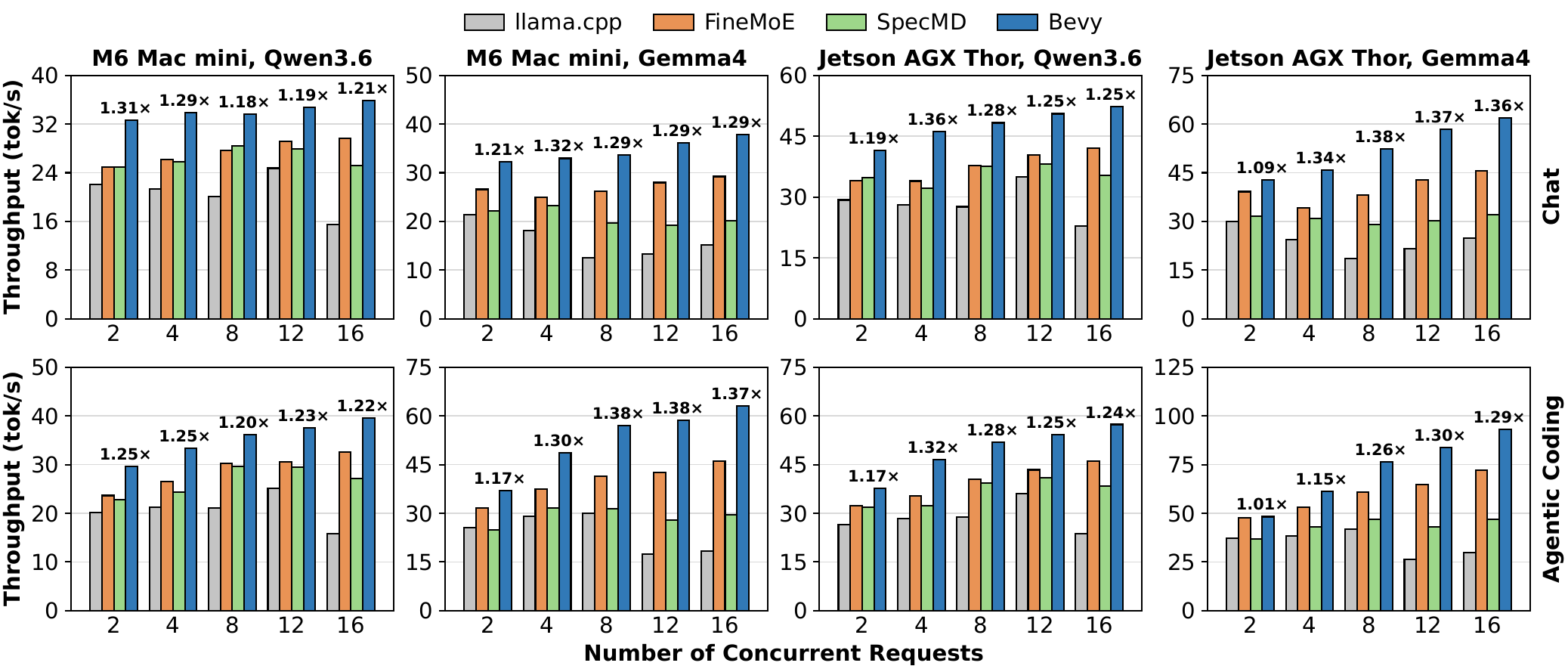}
\vspace{-5mm}
\caption{Throughput as the number of concurrent requests grows. The numbers above the bars give \sysname's speedup over the fastest baseline. Even without prefetching, \sysname achieves the highest throughput in every configuration.}
\label{fig:eval-concurrency}
\end{figure*}

\subsection{Experimental Setup}
\label{sec:eval-setup}

\noindent\textbf{Implementation.}
\sysname is implemented on llama.cpp~\cite{llamacpp} with support for Metal and CUDA. To support opportunistic layerwise grouping at runtime, we modify its graph builder to generate ggml computation graphs for individual jobs and retain intermediate states across jobs. The jobs in a group run as one batch in a single graph. We reuse compatible job graphs and captured CUDA graphs to reduce graph construction and launch overhead. \sysname uses LFU eviction rather than a recency-based policy, since recency alone can be a poor guide to expert reuse across layers~\cite{moeinfinity}. \sysname stores the experts on SSD in the expert-pack layout of SSD-LLaMA~\cite{ssdllama} and fetches them with eight reader threads.

\noindent\textbf{Testbed.}
We evaluate \sysname on two unified-memory platforms for local LLM inference: an Apple Mac mini with the M6 chip, 24\,GB of unified memory, and a 512\,GB SSD, and an NVIDIA Jetson AGX Thor Developer Kit with 128\,GB of unified memory and a Samsung 9100 PRO 2\,TB SSD (PCIe~5.0~$\times$4).

\noindent\textbf{Models and workloads.}
We use two MoE models, Qwen3.6-35B-A3B~\cite{qwen36} and Gemma4-26B-A4B~\cite{gemma4}, both quantized to Q4\_0. We run two workloads, Chat and Agentic Coding, with requests drawn from ShareGPT~\cite{sharegpt} and SWE-agent trajectories~\cite{sweagenttraj}, respectively.

\noindent\textbf{Baselines.}
We compare \sysname with three baselines, all implemented in the same engine. \textbf{llama.cpp}~\cite{llamacpp} uses its built-in GPU expert cache with LRU eviction and does not prefetch. \textbf{SpecMD}~\cite{specmd} evicts experts with its Least-Stale policy. Like Mixtral-Offloading~\cite{mixtraloffloading}, it applies the next layer's gate to the current layer's gate input to predict which experts to prefetch. We set it to prefetch each token's top-$k$ experts with an overfetch factor of 1.5 (12 experts per token). \textbf{FineMoE}~\cite{finemoe} prefetches the experts predicted by the most similar of the expert maps recorded from past tokens, and evicts the expert with the lowest product of usage frequency and predicted probability. We port FineMoE's open-source implementation as is, with an expert map store of 4096 maps, the largest the implementation allows, and a prefetch distance of 3 layers. All baselines fetch missing experts from SSD in the same expert-pack layout and with the same eight reader threads as \sysname. Because prefetches share these readers with the fetches of missing experts, FineMoE and SpecMD use at most two of them for prefetching at a time, so that prefetches delay those fetches as little as possible.

\noindent\textbf{Measurements.}
We report the decode throughput (tok/s) as the main evaluation metric. Each run decodes 512 steps, and we exclude the first 256 steps, which warm up the expert cache. We also exclude steps with observed SSD-read slowdowns from all systems in each comparison. On Thor, we fix the CPU and GPU clocks at their maximum with \texttt{jetson\_clocks}. On M6, the GPU clock drops whenever the GPU idles in the baselines, slowing them down considerably. To exclude this clock effect from the comparison, the baselines run a dummy kernel whenever the GPU would otherwise idle, which keeps the clock from dropping.

\subsection{End-to-End Performance}
\label{sec:eval-e2e}

We first compare the decode throughput of \sysname and the baselines as the number of concurrent requests and the expert cache size vary.

\subsubsection{Number of Concurrent Requests}
\label{sec:eval-concurrency}

Figure~\ref{fig:eval-concurrency} shows the decode throughput as the number of concurrent requests grows from 2 to 16, with 30\% of the experts cached. We measure five concurrency levels for both workloads, both models, and both devices. \sysname achieves the highest throughput in all 40 configurations. It is 1.01--1.38$\times$ faster than the fastest baseline and 1.29--3.46$\times$ faster than llama.cpp. The prefetching baselines are faster than llama.cpp, which does not prefetch, in almost all configurations, and the faster of the two is 1.13--2.52$\times$ faster than llama.cpp. With four or more concurrent requests, \sysname's speedup over the fastest baseline is 1.15--1.38$\times$ on both devices, for both models and both workloads. The speedup is smallest with two requests in most configurations, down to 1.01$\times$ for Gemma4-26B-A4B on Thor, since only one other request can run ahead while a request waits for its experts. Even without prefetching, \sysname is faster than every baseline at every concurrency level.

\subsubsection{Expert Cache Size}
\label{sec:eval-cache}

\begin{figure}[t]
\centering
\includegraphics[width=\columnwidth]{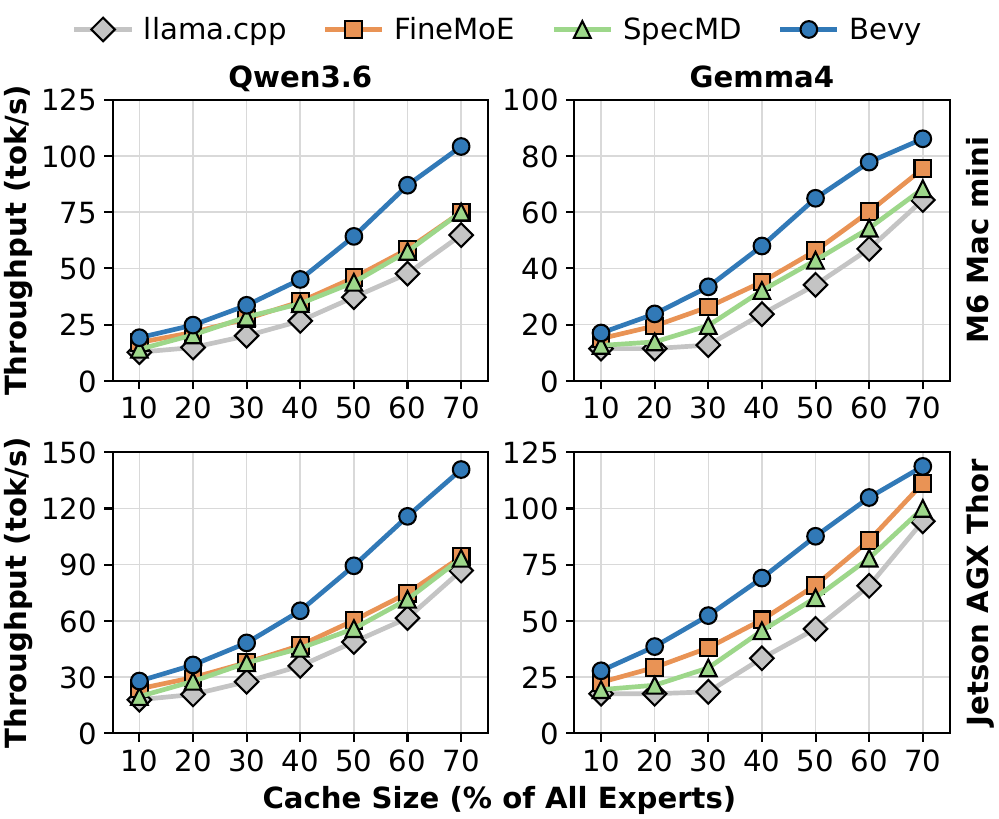}
\vspace{-5mm}
\caption{Throughput across expert cache sizes. As the cache grows, prefetch precision drops and the prefetching baselines approach llama.cpp, while \sysname achieves the highest throughput at every cache size.}
\label{fig:eval-cache-size}
\end{figure}

Figure~\ref{fig:eval-cache-size} shows the throughput as the expert cache grows from 10\% to 70\% of all experts, with eight concurrent Chat requests. \sysname achieves the highest throughput at every cache size, 1.07--1.55$\times$ faster than the fastest baseline and 1.26--2.83$\times$ faster than llama.cpp. As the cache grows, the prefetching baselines approach llama.cpp. FineMoE's speedup over llama.cpp reaches 1.43--2.06$\times$ at 20--30\% of the experts and drops to 1.08--1.18$\times$ at 70\%, as prefetch precision decreases with a larger cache (Figure~\ref{fig:prefetch-goodput}). In contrast, while \sysname is bound by the link, fewer missing experts shorten its link time, and its throughput increases accordingly. In this regime, \sysname's step time exceeds its link busy time by only 4--9\,ms on Thor, up to 40\% of the experts for Gemma4-26B-A4B and 60\% for Qwen3.6-35B-A3B. \sysname's speedup over the fastest baseline reaches 1.40--1.55$\times$ at 50--60\% of the experts on M6 with both models and on Thor with Qwen3.6-35B-A3B.

\subsection{Where the Speedup Comes From}
\label{sec:eval-breakdown}

\begin{figure}[t]
\centering
\includegraphics[width=\columnwidth]{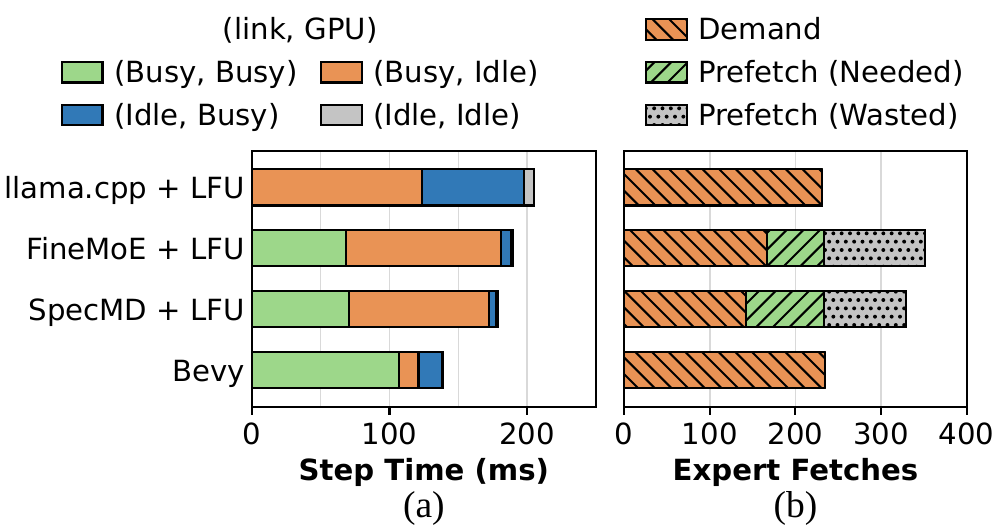}
\vspace{-5mm}
\caption{(a) Step time split by whether the link and the GPU are busy or idle. (b) Expert fetches per step: missing experts fetched on demand, and prefetched experts that the step needed or wasted. \sysname keeps the link and the GPU busy together for most of the step while fetching only the experts that the requests need.}
\label{fig:eval-gpu-fetch}
\end{figure}

To understand where \sysname's speedup comes from, we break down the step time and the expert fetches of each system for Gemma4-26B-A4B with eight Agentic Coding requests and 30\% of the experts cached on M6. To exclude the effect of the eviction policy, we replace the eviction policy of each baseline with LFU, which \sysname uses, in this experiment.

Figure~\ref{fig:eval-gpu-fetch}(a) shows how much computation and fetching overlap. In llama.cpp + LFU, the link and the GPU are never busy at the same time, and the GPU is idle for 60\% of each step while the link fetches. FineMoE and SpecMD overlap fetching with computation for 36--40\% of the step, but the GPU still waits for the link for 57--59\% of the step. \sysname keeps the link and the GPU busy together for 77\% of the step, and the GPU waits for the link for only 10\%. Although \sysname's GPU busy time is 1.6--1.7$\times$ that of the baselines, \sysname overlaps 86\% of it with fetching, so its step time is shorter.

Figure~\ref{fig:eval-gpu-fetch}(b) shows that the prefetching baselines fetch many experts that are not needed. FineMoE and SpecMD fetch 1.4--1.5$\times$ as many experts per step as llama.cpp + LFU, and 51--64\% of the experts they prefetch go unused. This agrees with the simulation in \S\ref{sec:bm-motivation} under the same LFU policy. \sysname fetches 235 experts per step, almost the same as llama.cpp + LFU. Together, these results answer the question raised in \S\ref{sec:bm-motivation}: \sysname reduces idle link time through request scheduling while fetching only experts known to be needed.

\subsection{Ablation Study}
\label{sec:eval-ablation}

\begin{figure}[t]
\centering
\includegraphics[width=\columnwidth]{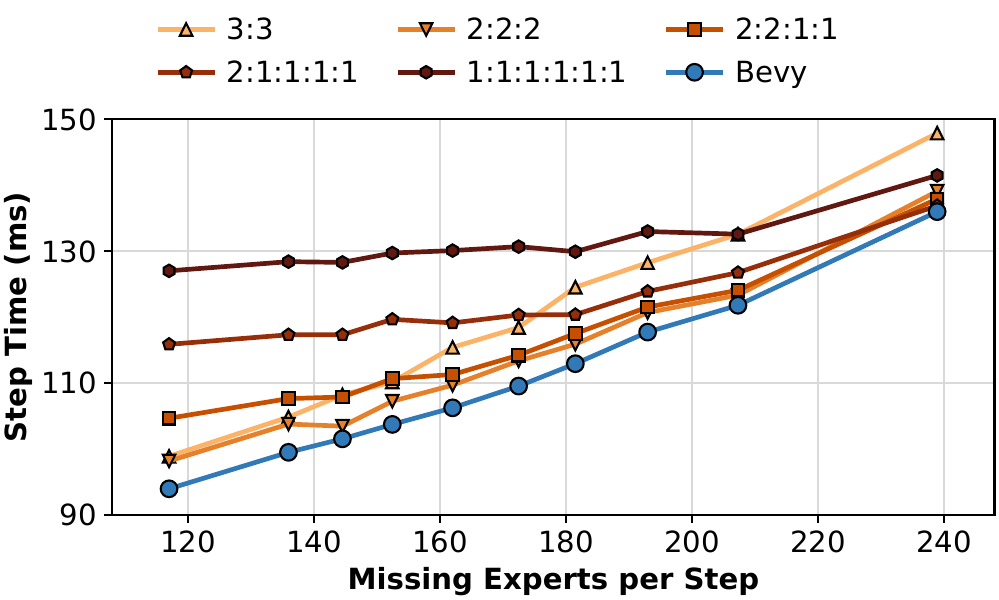}
\vspace{-5mm}
\caption{Mean step time against the number of missing experts per step, for static groupings and \sysname's opportunistic layerwise grouping. Which static grouping is fastest depends on the number of missing experts, while \sysname achieves the lowest step time at every point by adapting groups to the missing experts at each layer.}
\label{fig:eval-merge-states}
\end{figure}

We next evaluate \sysname's design components individually.

\subsubsection{Opportunistic Layerwise Grouping}
\label{sec:eval-batching}

Using the same setting as Figure~\ref{fig:c1} with six concurrent requests, we compare \sysname's opportunistic layerwise grouping with static groupings, which split the requests into fixed groups in advance and schedule each group as one unit: 3:3, 2:2:2, 2:2:1:1, 2:1:1:1:1, and 1:1:1:1:1:1. Figure~\ref{fig:eval-merge-states} shows that which static grouping is fastest depends on the number of missing experts. \sysname achieves the lowest step time at every point, 0.9--4.3\,ms below the fastest static grouping at that point, because it dynamically adapts the grouping granularity to the missing experts at each layer. This result shows that the self-balancing property described in \S\ref{sec:design-merge} works effectively.

\subsubsection{Lowest-Layer-First Scheduling}
\label{sec:eval-priority}

\begin{figure}[t]
\centering
\includegraphics[width=\columnwidth]{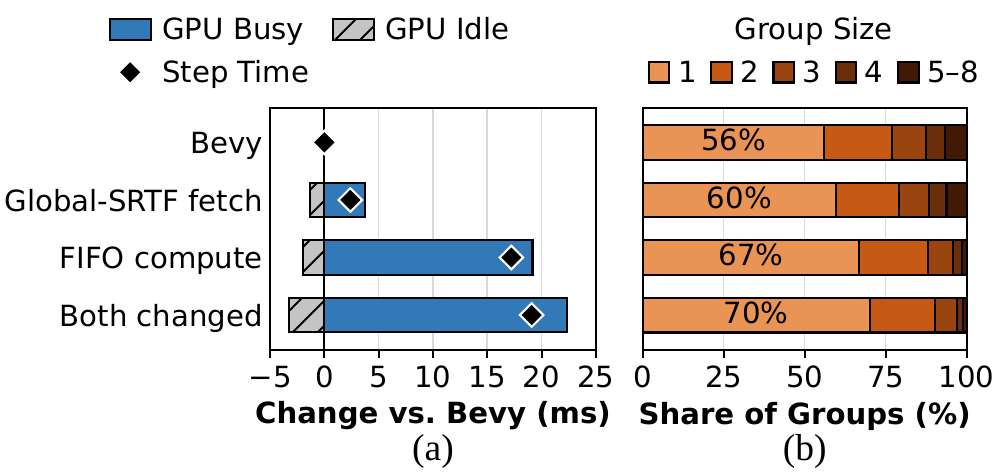}
\caption{(a) Change in GPU busy time, GPU idle time, and step time relative to \sysname when the fetch order, the compute order, or both are changed. (b) Share of groups by group size, the number of jobs that run together as one batch. Lowest-layer-first compute ordering creates batching opportunities: it forms larger groups, and the GPU busy time it saves outweighs the increase in GPU idle time.}
\label{fig:eval-order}
\end{figure}

To evaluate the lowest-layer-first compute and fetch orders of \S\ref{sec:design-priority}, we replace each of them in \sysname with an order that ignores request progress, while keeping the grouping rule of \S\ref{sec:design-merge}. For computation, FIFO over ready jobs selects the layer of the oldest ready job instead of the lowest layer. For fetching, global SRTF fetches the missing experts of the job with the fewest remaining missing experts, regardless of its layer. We also replace both orders. We use the same setting as Figure~\ref{fig:eval-gpu-fetch}.

Figure~\ref{fig:eval-order} shows that the compute order has the larger effect. With FIFO, the step is 12.3\% longer because the jobs run in smaller groups, which repeats reads of shared weights and increases compute time (\S\ref{sec:idea-challenges}). The share of groups with a single job rises from 56\% to 67\%, and the share of groups with three or more jobs falls from 23\% to 12\%. The GPU busy time grows by 19.2\,ms, while the GPU idle time shrinks by 2.0\,ms. The fetch order has a smaller effect: with global SRTF, the step is 1.7\% longer, as the share of groups with a single job rises to 60\% and the GPU busy time grows by 3.7\,ms. Replacing both orders makes the step 13.6\% longer. In all three cases, the GPU idle time does not increase, so the steps are longer because of additional computation rather than waiting. This confirms that lowest-layer-first effectively creates batching opportunities, as described in \S\ref{sec:design-priority}.

\subsection{Combining \sysname with Prefetching}
\label{sec:eval-prefetch}

\begin{figure}[t]
\centering
\includegraphics[width=\columnwidth]{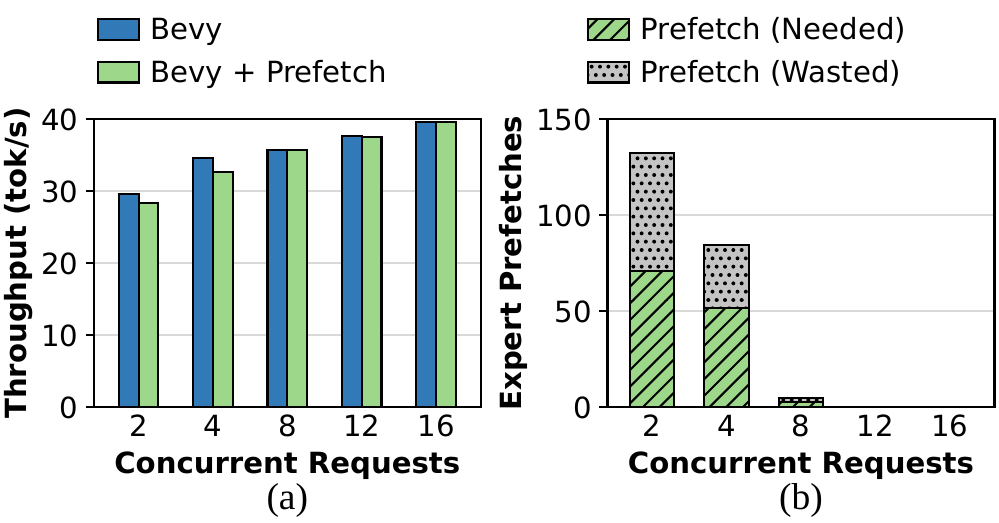}

\caption{(a) Throughput of \sysname and \sysname + Prefetch as the number of concurrent requests grows. (b) Expert prefetches per step, split into needed and wasted. Prefetching does not improve the throughput of \sysname: with few requests, many prefetches are wasted, and with more requests, the link has no idle time left for prefetching.}
\label{fig:eval-prefetch}
\end{figure}

\sysname fetches experts only on demand, and prefetching can also be applied on top of it. To evaluate the effect of combining \sysname with prefetching, we apply the prediction method of Mixtral-Offloading~\cite{mixtraloffloading} to \sysname (\sysname + Prefetch). While job~$l$ of a request runs on the GPU and the link is idle, \sysname fetches the request's predicted experts of layer~$l{+}1$ that are not in the cache. The prefetch mechanism is the same as in the baselines (\S\ref{sec:eval-setup}). We use Qwen3.6-35B-A3B with Agentic Coding requests and 30\% of the experts cached on M6, and vary the number of concurrent requests from 2 to 16.

Figure~\ref{fig:eval-prefetch}(a) shows that prefetching does not improve the throughput of \sysname. With two and four requests, \sysname + Prefetch has 4.2\% and 5.4\% lower throughput. There are two reasons. First, even when a prefetched expert is needed, it is almost always still being fetched when the next job needs it, so it saves little time. Second, as Figure~\ref{fig:eval-prefetch}(b) shows, 46\% and 39\% of the prefetched experts are wasted, and these wasted fetches only delay the fetches of missing experts. From eight requests on, missing experts wait to be fetched for almost the whole step, so the link has no idle time left for prefetching. As a result, \sysname + Prefetch rarely prefetches and performs the same as \sysname. These results show that adding prefetching with the prediction method of Mixtral-Offloading does not improve the throughput of \sysname.

\section{Related Work}
\label{sec:related}
\noindent\textbf{Overlapping computation with expert fetches.}
Expert offloading systems such as Klotski~\cite{klotski} and ExpertFlow~\cite{expertflowshen} compute the experts already in GPU memory first while the missing experts are fetched. However, all requests in a batch still move to the next layer together, only after all the experts the batch needs at the current layer are fetched. To remove this barrier, KTransformers~\cite{ktransformers} uses Expert Deferral, which defers the computation of some experts and proceeds to the next layer. This, however, is an approximation that changes the model output. \sysname instead lets different requests run at different layers to overlap computation with expert fetches, and does not change the model output.

\noindent\textbf{Request batching for expert sharing.}
To reduce the expert fetches of a batch, other systems choose which requests to batch together so that the requests in a batch activate many of the same experts~\cite{expertflow,saber}. \sysname addresses a different question, how to schedule those requests within a single step to overlap computation with expert fetches. Since the two approaches are orthogonal, \sysname can schedule the requests that these systems batch together.

\noindent\textbf{Expert parallelism in large-scale MoE serving.}
In large-scale server inference, expert parallelism distributes the experts across many GPUs that exchange tokens with each other. To hide this communication behind computation, prior systems split a batch into micro-batches and pipeline GPU computation with communication~\cite{megascaleinfer,deepseekv3}. \sysname differs from them in both its goal and its method. Within the limited memory of a local device, its goal is to keep the GPU from stalling on misses while making full use of the link bandwidth. Rather than splitting a batch into micro-batches, \sysname groups jobs dynamically according to the progress of each request (\S\ref{sec:design-merge}).

\noindent\textbf{CPU--GPU hybrid systems.}
In local MoE inference, where not all experts fit in GPU memory, many systems keep the experts in CPU memory and use the compute resources of the CPU alongside the GPU~\cite{ktransformers,freetoken,wang2026,fiddler}. These systems, however, cannot be applied to devices with a unified memory architecture, such as Apple silicon and other systems-on-chip (SoCs) with integrated GPUs, which are common for running LLMs locally. \sysname is implemented on unified memory, but it also applies to discrete GPUs by offloading the experts to CPU memory instead of SSD. Extending \sysname to use CPU computation as well is possible, and we leave it for future work.

\section{Conclusion}
\label{sec:conclusion}

Our analysis showed that expert caching reduces prefetch precision in local MoE inference, wasting bandwidth on unneeded experts.
\sysname uses runahead inference to let ready requests reach later layers and reveal actual expert demand early.
It fetches the resulting missing experts while computing other requests, without predicting expert selections.
Its self-balancing scheduler batches ready requests by layer and prioritizes lagging ones, preserving shared-weight reuse.
Across two local platforms, \sysname achieved 1.26--3.46$\times$ the decode throughput of llama.cpp and 1.01--1.55$\times$ that of the fastest prefetching baseline.

\bibliography{references}
\bibliographystyle{mlsys2025}

\end{document}